# Review of Barriers for Federated Identity Adoption for Users and Organizations

John Sherlock, Manoj Muniswamaiah, Lauren Clarke, Shawn Cicoria
Seidenberg School of Computer Science and Information Systems
Pace University
White Plains, NY, US
{js20454w, mm42526w, lc18948w}@pace.edu, shawn@cicoria.com

*Abstract*— A look at Identity as a Service (IDaaS) and Federated Identity Management (FIM) and acceptance amongst organizations, users, and general population. While FIM has shown acceptance amongst educational, commercial and government organizations, the general population acting has not seen the level of trust as the former. What are the barriers or enablers for acceptance that might allow, in the extreme example, the ability to logon to a bank with your Facebook credentials and transact business?

## I. Introduction

Many solutions, products, and protocols exist for establishing federated identity for enabling users to securely logon to Service Providers (SP) for various purposes, yet, there is still a chasm with regards to trust. That is whether consumers trust both Identity Providers (IdP) and SP to participate in a federation.

We take a look at research that examines commercial and public organizations and various implementation impediments and facilitators. Additionally, while Federation amongst universities, government, and, commercial partners is much more prevalent, we lack that level of acceptance for the general population. We take a look at government efforts, most notably in the European Community towards establishing a common credential provider and broker-based approach.

Finally, for future research, we ask a rhetorical question – when will I be able to use my Facebook Identity to logon to my bank?

## II. Federation identity basics

Initially organizations started to use Single-Sign-On (SSO) to unify authentication systems together for better management and security. SSO was adopted across different internal applications and via APIs for external applications. With platforms spread across different devices bringing in together user authentication and authorization with increased security is a challenge which Federated identity tries to address [5].

In Federated identity an application uses an identity management system that stores user's electronic identity to authenticate. It decouples authentication and authorization functions. It avoids a situation where every application has to maintain a set of credentials for every user [5].

## III. Government Sponsored Federation

Since 1995, EU member nations have been implementing various forms of electronic Identity Management (eIDM) based on the interoperability of eSignatures. The lack of an EU-wide legal framework governing such an implementation has resulted in difficulties in defining actor responsibilities and liabilities. Cross-border requirements amongst EU countries have created incremental challenges. This has unfortunately resulted in a degree of "subsidiarity" amongst the member nations: each member nation maintaining its autonomy and responsibility. [1]

Both the U.S. and the U.K. governments have decided that citizens will authenticate to government resources using Federated Identity; however, neither government wishes to perform the role of IdP[1]. This defers to private industry the opportunity to offer identity management solutions to consumers while still not answering questions such as (1) who will oversee the solution design and implementation, and (2) who will control and monitor the governance of the solution. What limited functionality has been implemented on the consumer side in the UK has been received with a reasonable level of trust on the part of the consumer, with ease-of-use scoring the highest. Opportunities have been identified[1]:
- 50% left the site once reaching the hub page,
- 25% left the site once reaching the consent page,
- 34% of users felt threatened, not reassured, by the privacy.

## IV. InCommon – Where it's working

Within the U.S. driven by the research community, "The InCommon Federation is the identity management federation for US research and education, and their sponsored partners" [2]. They currently service over 7.5 Million users with this federation amongst 430 education and nearly 200 other organizations[2].

The success of InCommon may be linked back to the Shibboleth model that was implemented, along with Trust in Privacy from the IdPs and alleviation of significant identity management functionality for SP[3]. This also was a step-up for many organizations in having better security than previous implementations. For example, JSTOR had been using IP address blocks to permit access from users cross Universities. The implementation of InCommon using Shibboleth allowed clarity in who is requiring access with some level of assurance that the actual user has been authenticated and permissioned for access by the supplying IdP amongst the trust.

## V. Consumer Federation

What may be core to the success of InCommon is a captive community of practitioners that utilize the services offered within the Federation for their daily needs. It may also be attributed to a community well versed in the underlying technology, or at least comfortable with it in general. We still haven't seen the ability to access your Bank Account through this federation.

Amongst the general population we have different motivations that may attribute to the slow uptake or resistance. In the UK, the Identity Assurance (IDA) program issued guidelines and mandates for implementation of an approach that has yet to take hold[1].

In a series of studies[1] various obstacles were cited ranging from poor User Interface Design, to Security and Trust concerns. Major concerns still exist over unauthorized release of privacy information, along with inability to understand associations of an IdP to a SP – one stating "PayPal have nothing to do with the National Health Service (NHS)"[1].

The lack of general Trust in the federation amongst the general public may be attributed to their naivety or lack of understanding of the overall Federation model. The same study suggests that this may be improved through better User Experience (UX) and also presents as part of that argument which examples of UX worked better or worse in their studies[1].

The same study also identifies which types of IdPs seem more trustworthy, or may be a candidate for use within a Federation scenario amongst the general population, with the post-office and cable providers being amongst the highest potential for "Would use" in a Federation scenario[1].

Establishing eIDM trust in the eyes of the general population has presented itself as a significant barrier to full scale end-to-end implementation of Federated Identity. These roadblocks include issues such as:
- No clear definition on what constitutes an eID,
- Disagreement on which governing body (governmental or private) should create and manage the eID,
- Disagreement on which governing body should manage and/or regulate the data flows amongst the various FIM SP,
- How the integrity/reliability/privacy of the identity will be ensured; who will stand behind this guarantee.

## VI. Risks

Trust, security and liability are three key areas of risk which are constantly challenged with FIM/IDaaS systems. The trustworthiness of the user, the IdP or SP may be in question by the entity who did not perform the identity screening. The development of the NIST[7] identity credentialing via the service levels is one method to assist SP and IdP in properly screening their members.

From a security perspective, Man-in-the-Middle attacks in an open network can lead to theft on identity credentials which allow access to confidential information. Misuse of the IdP and SP with user identity information causing the risk of IdP becoming malicious and risking the use of user identity information in SPs to share it with other SPs or third parties.[6]

In many of the types networks where FIM/IDaaS systems have been tried the questions always arises, what happens if something goes wrong? Or more importantly, who is liable? A system could become unavailable to authenticate users, causing problems for SP relying on its operation. And authentication itself could fail, allowing unauthorized users to be incorrectly authenticated as other users. In both cases, rules that determine which party is responsible are a potential source of conflict.[3] Liability arrangements must be clearly defined and articulated for all stakeholders.[3]

## VII. Summary and further research

Globally, there is consensus that new research would be needed to coordinate existing knowledge and know-how into a coherent vision capable of being seamlessly implemented cross-border and cross-culture, spawning the required trust and confidence at all levels. [4]

A number of conceptual models exist which demonstrate that there is a need for a unified solution to federated identity management: for individuals, businesses and government agencies alike. These technological solutions have been proven to be viable and cost effective for many of today's e-commerce environments, yet full scale implementation is constrained by a low level of trust on the part of the general population of the technology and of the SPs. Partly the result of misinformation and misunderstanding, many of the FIM participants would be more inclined to participate in these eIDM offerings if they could only get beyond the trust issues.

Further research would be required to address consumer concerns:
- Who should 'own' (i.e. regulate) the solution?
- What would it take to raise confidence in any solution?
- What is the model for user recourse if the solution breaks?

Reviews of current state of eID implementations indicate that, at least for the short term, consumers may not be able to use a common access/authentication method for secure online resource access control.